\documentclass[twocolumn,showpacs,preprintnumbers,amsmath,amssymb]{revtex4}

\usepackage{graphicx}
\usepackage{dcolumn}
\usepackage{bm}

\begin{document}


\title{Investigation by physical methods of the possible role of telomeres in DNA\\in aging process}

\author{\large{Md. Ashrafuzzaman and Ahmed Shafee$^{*}$ }}
\affiliation{%
Institut de Physique, Universit\'e de Neuch\^atel, Rue A.L.Breguet, 2000 Neuch\^atel, Switzerland\\
*Department of Physics, University of Dhaka, Bangladesh\\
contact: md.ashrafuzzaman@unine.ch }%


\begin{abstract}
The interaction energies between the different types of bases of a single strand of DNA molecule have been calculated. Using these original values of energies the harmonic behavior of  a number of  base patterns of DNA has been studied. In view of the great interest aroused by the discovery of the role of the telomere segment of the DNA in the replication process and its possible link with the aging process, we have investigated, with simple models, the harmonic behavior of the telomeric pattern of bases as well as the thermodynamic response in the biological system. With these results a conclusion on the probable role of the telomeric pattern on aging has also been drawn. Here the calculated values of harmonic frequencies of the telomeric pattern of bases and of other possible patterns show that the telomeric pattern is associated with the highest vibrational frequency among all patterns of base combinations at the replication end of DNA. This seems to suggest that due to  the existing telomeric pattern being closest to the frequencies of the electromagnetic radiation coming from sunlight, resonance of the telomeric frequency with  such radiation may be responsible for damage to the reproductive ability of the cells and consequent aging and other problems. On the other hand in the last part of this work we have calculated the thermal vibrational amplitudes of the telomeric pattern and other possible patterns which show that the amplitude for the telomeric pattern is the least, and this suggests that the telomeric pattern is more mechanically and thermally stable than other possible patterns in the biological environment against damage from ordinary heat and mechanical effects.
\end{abstract}

\maketitle

\section{ Introduction }

Most of the activities of the different parts of a living system are controlled in nucleic acids DNA (deoxyribonucleic acids) and RNA (ribonucleic acids).
The nucleic acids are made of some combinations  of purine and pyrimidine bases called genetic codons. The behavior and functioning of a living system are
all controlled by different combinations of genetic codons. The possibility of the control of genetic characteristics is the most important area of the
latest genome research. People are trying to find out how the genetic codes arrange themselves to make DNA and how they have evolved with time. Since the
central dogma of biology is that  genetic changes are brought about by changes in the macromolecules, it is of great importance also to investigate the
stability of the molecules and their patterns responsible for the genotypical variations. It can of course be expected that some of the patterns of bases
are more stable and they may be more persistent in organisms.

Attempts are being made to gain a full understanding of the role of the different molecules involved in the passing of genetic information to succeeding
generations and of the pitfalls which lie in the path. That structure and function of macromolecules are interrelated is fully established at present.
Hence any effort to understand biological processes at the molecular level requires a reasonable understanding of their structure, which in turn, can only
be achieved by a thorough study of stability and energy considerations. Molecular biophysics, therefore, is an integral part of man's quest for complete
control of his inner world.

For a long time biochemists and biologists have been trying to gather more information on structural and functional behavior of biomolecules in terms of
enzymatic activities. They met some success in certain areas but the complete ideas of the behavior of the molecules are to be established. The nature of
biochemical or enzymatic approach to know the structural and functional behavior of biomolecules is biological aspects and the limitations of these methods
force the scientists to find other ways to explain the cause of stability or instability and some functional behavior of biomolecules. Therefore scientists
are trying to impose the physical methods to explain the structural and functional behavior of biomolecules. In this new process to learn the behavioral
aspects of biomolecules, the established methods or sometimes some modifications of the existing methods according to the problem of real or imaginary
structures of biomolecules and their functioning are used.

For more than two decades physicists have been studying the structure-function relationship of the proteins. People now have a fairly good idea about how
muscle filaments work or how some of the simpler enzymes perform their catalytic activities. Protein crystallography is an established science and using
this method coordinates of the atoms of the smaller proteins, such as dihydrofolate reductase (DHFR), are known accurately. Many research groups all over
the world are active in the ambitious task of designing molecules which will accelerate or inhibit the action of enzymes by attaching themselves to the
enzyme at the active site. While this approach does remain a promising field, more recently it is becoming attractive to try to do the same with nucleic
acids, i.e., to suppress or promote specific genes. Genes that cause diseases should be repaired, and if that is not possible, at least those causing
production of harmful proteins should be inhibited. Some antibiotics prevent bacterial growth by stopping nucleic acid (NA) synthesis, others attempt to
arrest the growth of cancer cells. However present methods do not allow choice of specific targets, e.g., bacterial NA or the NA of cancer cells. More
knowledge is required about the conformational structures of the NAs, about their flexibility and their binding to different ligands. Here too energy
considerations are of vital importance.

Recently scientists have become aware of the existence of the link between aging and NAs. It is well known that cells can not replicate
indefinitely in vivo. It is possible that despite the high accuracy of the process of replication, with proofreading arrangements to correct
errors, it is impossible to eliminate errors altogether. The biological process continues independently and man can not control any biological
activity completely. Any disorder initiated inside the biological system therefore sometimes grows multi-dimensionally. The random noise is
accumulated and eventually replication stops.

Very recently scientists have begun to look into another possible source of loss of replicability of nucleic acids. The DNA molecule contains a
tail called telomere, a certain pattern of bases which do not code for any gene, but on which the DNA polymerase has to stand when performing
replication. It has been suspected that with each new cell generation in an organism, i.e., in the replication of somatic cells, parts of the
telomere gets lost. As the telomere keeps shrinking with age, eventually a time comes when they can no longer replicate. As dying cells are not
replaced by new cells, the organism decays or 'gets old' and finally faces death. This biological activity of DNA [1,2] is known to scientists
but they don't know why and how this activity occurs.

M. Olovnikov formulated the problem of terminal under replication of linear DNA molecules in 1971[3]; this phenomenon is caused by the inability
of DNA polymerases to replicate several nucleotides at 3´ ends of DNA templates. Olovnikov also suggested that a specific biological mechanism
should normally prevent this phenomenon. This mechanism was expected to be active in gametes, cancer cells, as well as in cells of vegetatively
reproducing organisms. In most other cases, e.g., in many human somatic cells, this mechanism is suppressed.

Further studies revealed the enzyme telomerase [4,5] (whose existence had been predicted by Olovnikov) that compensates for DNA shortening in the
mentioned cell types. The function of telomerase is to add a repeated sequence (the hexamer TTAGGG in humans), which forms the so-called telomere,
to ends of nuclear DNA. Here A is for adenine, G for guanine and T for thymine. After this, underreplication of the linear DNA molecule only
shortens this nontranscribed sequence of the telomeric fragment of the chromosome without damaging the genetic information or the mechanism
that reads it. At certain stages of development in early embryogenesis, the gene encoding telomerase in the majority of human somatic cells is
switched off, thereby making the genome susceptible to shortening. The telomere shortens at a low but appreciable rate which impairs the
functioning of the chromosome. This impairment begins long before the disappearance of the whole telomere, which removes protection from genetic
information contained in transcribed regions. This role is still poorly understood.

There is close correlation between shortening of telomeric DNA regions and Hayflick's limit [1,2]. To surpass this limit and continue
reproduction, the cell should activate its telomerase gene.

The most remarkable feature is that switching off the telomerase gene is an ontogenetic stage that occurs at a distinct time point in the life of
an organism and involves only some of its cell types. This event seems to perform a specific function and cannot be regarded as a disorder in the
living system or a kind of unpredictable age-related defect, although it clearly promotes aging. In this context, we are reminded of an
observation made in experiments with barley germs. During the development of the germ, the telomere suddenly loses 50 kb (kilo bases). It loses
an additional 20 kb during growth of the spike [6]. The mechanism responsible for this event remains unknown. If the arrest of telomerase
synthesis is considered as an accidental fault, telomere shortening in cells containing no functional telomerase appears to be an act of
deliberate damage to the organism. An alternative interpretation is that the inhibition of the telomerase gene and the telomere shortening are
biologically important events. To understand the meaning of this process, one should remember that telomerase genes of somatic cells are
switched off only in sexually reproducing organisms, but not in vegetatively reproducing organisms. It is exactly in the former case that the
appearance of a new trait, which can result from a combination of parental genomes, becomes the most probable event. However, in vegetatively
reproducing organisms, the appearance of a new trait results from random mutations occurring in the same cell.

The above mentioned properties of the telomere make understanding the cause of its instability is an exciting problem. Again, it is possible that
evolution has chosen the most stable pattern. This may be investigated by studying the energy of the telomere portion of the DNA. We can try to
interpret this stability with respect to radiation, vibration etc i.e., by using  physical rather than  biological methods. If the energies
involved between the bases of DNAs can be determined and if using these harmonic behaviors of the telomere can be calculated
(i.e., the harmonic frequencies and thermal vibrational amplitudes of bases) it can probably be possible to get some ideas about the stability
of this all important tail which needs to be retained for good health and to continue normal living behavior. Application of quantum mechanics
can give a very good solution of energies and other necessary parameters to explain the states of a single body or in some cases two or three
body problems. The absolutely correct way of carrying out such investigations would be to solve the Schrodinger equation for the entire DNA,
which is, of course, quite impossible. Even small molecules can not be solved quantum mechanically, exactly or even by approximation methods.
However, one can, as a zeroth order approximation, completely ignore the details of the interactions and study only the symmetry aspects of the
problem by taking square well potentials or harmonic behavior, as is often done in condensed matter physics. The harmonic attitude has been
adopted in this work and the telomere has been studied as a specific pattern of bases represented by harmonic oscillators. For this work the
bases are considered as simple atoms connected to the nearest atom (base) by a force constant '$k$' i.e., between the bases linear restoring
forces are involved. We believe that even with such a drastic approximation a meaningful beginning can be made in trying to understand the
stability problem of the telomere and the response of the pattern with the thermal, electromagnetic and other agents involved in environment.
We have also introduced the thermodynamics effect on the telomere molecules in this work which would represent the dependence of structure and
function of biological molecules, especially of telomere on thermal condition of the biological system.

In sec.II, the energies between the different bases in DNA have been calculated. For this some established methods of energy calculation have
been used and the values of some parameters needed for this work of calculating the energies involved in DNA have also been calculated. In sec.III,
a method of calculating the harmonic frequency of telomere and other possible patterns taking four purines and two pyrimidines and also for few
other imaginary patterns of bases has been developed. For this a matrix has been developed and using the energies involved between the bases in
DNA the harmonic frequency of telomere has been calculated and a conclusion on the stability of telomere pattern in the replication end of DNA
has been drawn. In sec.IV, the thermodynamic behavior of DNA molecules especially the temperature effect on the stability of telomere has been
discussed. Here it is tried to find the effect of the environment especially the biological environment where the DNA exists on the structure and
function of telomere. Finally in sec.V, the results are discussed and the conclusions are presented.

\section{ Energy calculation for Base Sequences in DNA}

\subsection{ Introduction}

In DNA the successive bases exist in energetically equilibrium state. The most important energies between the successive bases are stacking energy and van der Waals interaction. Stacking energy is due to interaction between the magnetic moments of the helical structure of bases and van der Waals interaction is the well known 6-12 potential that arises when the two bases come very near to each other. In this work these two types of interactions between the bases of DNA have been calculated using the well known formulae.

\subsection{ Energy Modeling}

A computer model has been constructed to make DNA double helices according to whatever base pattern sequence we required. This was done by
smoothly fitting experimental data for each base pair configuration with one another to form a straight double helical structure . However some
flexibility were also retained so as to vary the inter base separation to some extents. While lengthening the double helix by stretching it our
method also changes the twist angles to some extends so that the interbase angles which are normally $36^o$ for 10 base pairs per turn per piece
can vary to some extends in keeping with the inter base separation $h$ which is normally 0.36 nm [7-12].
It is well known that the most important energy in forming double helix is stacking energy. The model described here has been used to calculate this stacking energy for different base pair neighbors. The pyrimidine (Cytosine and Thymine ) have single aromatic six side a ring pattern whereas for the ring in purine (Adenine and Guanine) each has adjoining six sided and a five sided aromatic rings. These rings content  electrons that contribute magnetic moments.
The energy between two rings i.e., between two such magnetic moments is given by
\begin{equation}
U({\bf{r_1}},{\bf{r_2}})=\frac{\mu_o}{4\pi}\lbrack \frac{{\bf{\mu}}_1\cdot{\bf{\mu}}_2}{\vert{\bf{r}}_1-{\bf{r}}_2\vert ^3}-
3\frac{{\bf{\mu}}_1\cdot({\bf{r}}_1-{\bf{r}}_2) {\bf{\mu}}_2\cdot({\bf{r}}_1-{\bf{r}}_2)   }{\vert{\bf{r}}_1-{\bf{r}}_2\vert ^5}  \rbrack
\end{equation}
When taking into account the orthogonality of the s (i.e., the base planes) with the helical axis, this reduces to
\begin{equation}
U(\vert{\bf{r_1}}-{\bf{r_2}}\vert)=\frac{\mu_o(lg)^2}{4\pi}\lbrack 1-3\frac{d^2}{s^2}\rbrack
\end{equation}
where $\mu_o=4\pi\times10^{-7}$ SI unit, is the magnetic permeability in free space, $s$ is the actual distance between the centers of the two rings and $d$ is the 2 dimensional projection of the separation in the plane of the base $l$ is a no. and $r$'s are coordinates. $\mu_1$ and  $\mu_2$ are proportional to the angular momentum of  electrons by the usual gyromagnetic ratio ($g$) and can be taken identical for other rings because always  $\pi$ electrons are involved.

It is to be mentioned the part of energy that can counter the DNA to be collapsed by the attractive stacking energy. This type of energy is the van der Waals interaction which is  given by
\begin{equation}
U(i,j)=-\frac{A}{r_{ij}^{6}}+\frac{B}{r_{ij}^{12}}
\end{equation}
where $A$ and $B$ are van der Waals constants. The van der Waals interactions becomes dominating when two atoms come within van der Waals limit distance. This type of interaction is also called the 6-12 potential.

Special feature for DNA double helix is that all the base rings are perpendicular to the axis of the double helix. This feature is included in the energy calculation. It has also been taken in the term the rotation of the bases as the helix goes down as is already mentioned.
Combining the two energies mentioned above the equation for the calculation of the total interaction energy [13] between the two bases of DNA is being calculated which is as follows
\begin{equation}
V_{ij}=\frac{A}{(z^2+r^2)^6}+\frac{B}{(z^2+r^2)^{3/2}}(1-3\frac{z^2}{z^2+r^2})
\end{equation}
where $z$ represents the variable separation between the $i$ th and $j$ th bases and $r$ represents the radius.

In the real configuration of DNA the average separation of two successive bases is about 0.36 nm and in the calculation of the total energy here this
separation has been used as the minimum energy state and thus this state is more stable. Using this minimum energy state at 0.36 nm separation the
constants $A$ and $B$ have been calculated. In calculating the values of the force constants the value of energy has been taken from the deviated portion
from this minimum level.

\subsection{ Results}

The results for the stacking energies and the total energies including the van der Waals energy with respect to the different distances between the bases
in DNA are shown in Fig.2 and Fig.3 respectively.  In these calculations the stacking energies between purine-purine bases at different separations are
seen to be higher towards negative than those between purine-pyrimidine and pyrimidine-pyrimidine bases respectively. Similar sequence is also seen for
the calculation of total interaction energies between the bases.

\subsection{ Discussion}

The stacking energy calculation between the bases shows that the stacking energies for different order of separation between purine purine bases are higher than those between purine pyrimidine and pyrimidine pyrimidine bases. This is because pyrimidine (Cytosine and Thymine ) have single aromatic six side a ring pattern whereas for the ring in purine (Adenine and Guanine) each has adjoining six sided and a five sided aromatic rings. So for purine purine interaction extra stacking for extra rings exists and therefore the stacking energy between purine purine bases is much higher than that between pyrimidine pyrimidine bases. When the two rings come close due to their attraction by stacking energy they should collapse as the stacking interaction is only attractive. So in lieu of being stable the whole DNA molecule should collapse. But this type of collapsing is encountered by van der Waals high repulsive 12 potential which becomes dominating when the charge distributions of two atoms overlap. So the attractive interaction of stacking energy is thus compensated by the high repulsive van der Waals interaction.

\section{ Frequency calculation for different Base sequences in DNA}

\subsection{Introduction}

DNA bases are considered as molecules those are attached to each other by intermolecular forces that are harmonic. So between two bases there
exists force constants. The values of force constants $k$ between different pairs e.g., purine-purine, pyrimidine-pyrimidine, purine-pyrimidine
etc. are different. Due to this force DNA molecules follow the harmonic behavior and therefore they vibrate with some harmonic frequencies.

\subsection{ Derivation of Expression for Frequency}

The values of $k$'s are calculated from the van der Waals interaction and stacking interaction [7,11,12] between bases of different types e.g.,
purine-purine, pyrimidine-pyrimidine, purine-pyrimidine etc. Using these values the harmonic frequencies of the telomere in DNA are calculated.
To illustrate the technique for obtaining the resonant frequencies and normal modes it is considered in detail a model based on a linear symmetrical
six atomic molecule. All these six atoms are considered to be on one straight line, the equilibrium distances apart being denoted by $b$. For
simplicity it is considered that only vibrations along the line of the molecule, and the actual complicated interatomic potential are approximated
by two spring of two different or similar force constants on both sides of an atom. There are six obvious coordinates marking the position of the
six atoms on the line. The atomic arrangement then repeats for the telomere in DNA like AGGGTT then again AGGGTT and so on upto a huge number of
repeatation. If the mean positions of the atoms are considered as $x_1$, $x_2$, $x_3$, $x_4$, $x_5$, $x_6$ then the positions of the atoms repeat
and thus the arrangement completes a cyclic arrangement. In telomere system it is assumed that there exists two types of atoms: four purine (one
Adenine and three Guanine) and two pyrimidine (two Thymine). It is true that Adenine and Guanine have about the same mass which is considered as
$M$ and the mass of pyrimidine (Thymine and Cytocine) is different from that of purine and let this be $m$. Both these $M$ and $m$ have been
calculated from the total atomic weights of their constituent atoms. In the arrangement there may exist the sequences like purine-purine,
purine-pyrimidine, pyrimidine-pyrimidine and the force constants for these arrangements are considered $k_1$, $k_2$, $k_3$ respectively. If a force
constant $k$ in between two atoms of mass $M$ each, the potential energy in this system is given by Potential energy
\begin{equation}
V_2=\frac{1}{2}k(x_2-x_1-b)^2
\end{equation}
For a three atoms system this potential energy becomes
\begin{equation}
V_3=\frac{1}{2}k(x_2-x_1-b)^2+\frac{1}{2}k(x_3-x_2-b)^2
\end{equation}
with the atomic arrangement as follows \\
$m(x_1)-k-M(x_2)-k-m(x_3)$\\

Similarly if we increase the number of atoms according to our real problem of telomere (AGGGTT) in DNA with six atoms the arrangement becomes like this\\

 $M(x_1)-k_1-M(x_2)-k1-M(x_3)-k1-M(x_4)-k2-m(x_5)-k3-m(x_6)-k2-M(x_1)$\\

For this system of atoms the total potential energy $V$ is
\begin{eqnarray}
V&=& +\frac{1}{2}k_1(x_2-x_1-b)^2+\frac{1}{2}k_1(x_3-x_2-b)^2{}
\nonumber\\
&& +\frac{1}{2}k_1(x_4-x_3-b)^2+\frac{1}{2}k_2(x_5-x_4-b)^2{}
\nonumber\\
&& +\frac{1}{2}k_3(x_6-x_5-b)^2+\frac{1}{2}k_2(x_1-x_6-b)^2
\end{eqnarray}

Now we introduce coordinates relative to the equilibrium positions:
\begin{equation}
y_i=x-x_{oi}
\end{equation}
where
\begin{equation}
x_{o2}-x_{o1}=b=x_{o3}-x_{o2},\quad etc.
\end{equation}

The potential energy then reduces to
\begin{eqnarray}
V&=& +\frac{1}{2}k_1(y_2-y_1)^2+\frac{1}{2}k_1(y_3-y_2)^2{}
\nonumber\\
&& +\frac{1}{2}k_1(y_4-y_3)^2+\frac{1}{2}k_2(y_5-y_4)^2{}
\nonumber\\
&& +\frac{1}{2}k_3(y_6-y_5)^2+\frac{1}{2}k_2(y_1-y_6)^2
\end{eqnarray}
Hence the $V$ matrix has the form,
\begin{equation}
V_m=\left(\begin{array}{cccccc}
k_1+k_2 & -k_1 & 0 & 0 & 0 & -k_2\\
-k_1 & 2k_1 & -k_1 & 0 & 0 & 0\\
0 & -k_1 & 2k_1 & -k_1 & 0 & 0\\
0 & 0 & -k_1 & k_1+k_2 & -k_2 & 0\\
0 & 0 & 0 & -k_2 & k_2+k_3 & -k_3\\
-k_2 & 0 & 0 & 0 & -k_3 & k_2+k_3\\
\end{array}\right)
\end{equation}
The kinetic energy $T$ has the form
\begin{equation}
T=\frac{1}{2}M(\dot{y}_1^2+\dot{y}_2^2+\dot{y}_3^2+\dot{y}_4^2)+\frac{1}{2}m(\dot{y}_5^2+\dot{y}_6^2)
\end{equation}
The T matrix is diagonal which is of the form,
\begin{equation}
T_m=\left(\begin{array}{cccccc}
M & 0 & 0 & 0 & 0 & 0\\
0 & M & 0 & 0 & 0 & 0\\
0 & 0 & M & 0 & 0 & 0\\
0 & 0 & 0 & M & 0 & 0\\
0 & 0 & 0 & 0 & m & 0\\
0 & 0 & 0 & 0 & 0 & m\\
\end{array}\right)
\end{equation}

Combining these two matrices, the secular equation appears as
\begin{equation}
   \vert V_m-\omega^2T_m\vert=0
\end{equation}
or,
\begin{equation}
\left(\begin{array}{cccccc}
\frac{k_1+k_2}{M} & \frac{-k_1}{M} & 0 & 0 & 0 & \frac{-k_2}{M}\\
\frac{-k_1}{M} & \frac{2k_1}{M} & \frac{-k_1}{M} & 0 & 0 & 0\\
0 & \frac{-k_1}{M} & \frac{2k_1}{M} & \frac{-k_1}{M} & 0 & 0\\
0 & 0 & \frac{-k_1}{M} & \frac{k_1+k_2}{M} & \frac{-k_2}{M} & 0\\
0 & 0 & 0 & \frac{-k_2}{m} & \frac{k_2+k_3}{m} & \frac{-k_3}{m}\\
\frac{-k_2}{m} & 0 & 0 & 0 & \frac{-k_3}{m} & \frac{k_2+k_3}{m}\\
\end{array}\right)=0
\end{equation}

Eigen values of this matrix give the square of the frequencies($i\omega$). Here the values for $k$'s (at the separation 0.36 nm between bases) are used from the calculated values from the total energies (stacking and van der Waals interaction) between the bases of DNA. The masses of purine and pyrimidine are the total mass of their constituent atoms.

\subsection{Results}

The eigenvalues of the derived matrix gives the square of the frequencies ($i\omega$). Hence negative values of the eigen values give the real parts of the frequencies and positive values give the imaginary parts of the frequencies.
From the eigenvalues of the matrix the values of the frequencies of the harmonic oscillator of telomeric pattern have been calculated. Similarly rearranging the patterns of the matrices for other patterns of base combination of telomere and for some imaginary patterns taking all purine or all pyrimidines the eigenvalues and thus the frequencies have been calculated.

The highest values of the frequencies [13] of telomere and other different patterns of base combinations are as follows:\\

Telomere (AGGGTT) pattern:\\
$\omega$ =-3.104028110062502$\times 10^{12}$ Hz\\
$\omega$ =+3.104028110062502$\times 10^{12}$ Hz\\

AGTGGT pattern:\\
$\omega$ =-2.586737456962528$\times 10^12$ Hz\\
$\omega$ =+2.586737456962528$\times 10^{12}$ Hz\\

AGTGTG pattern:\\
$\omega$ =-2.944491229994985$\times 10^{12}$ Hz\\
$\omega$ =+2.944491229994985$\times 10^{12}$ Hz\\

AAAAAA or GGGGGG pattern:\\
$\omega$ =-3.333518830239974$\times 10^{12}$ Hz\\
$\omega$ =+3.333518830239974$\times 10^{12}$ Hz\\

TTTTTT or CCCCCC pattern:\\
$\omega$ =-1.668814797430148$\times 10^{12}$ Hz\\
$\omega$ =+1.668814797430148$\times 10^{12}$ Hz\\

\subsection{Discussion}

In the calculation of the frequencies of the oscillators of the telomere pattern of bases and the other patterns it is seen that the value of the frequency for the telomere pattern is the highest among those patterns having four purine and two pyrimidine bases. So the frequency of the telomere pattern is close to the frequencies available in the sunlight. So a resonance of this frequency with the frequencies of the infrared rays existing in the lower side of the band of light coming from the sun may occur. Most of the people stay under shade most of the time of their life. The infrared light of the frequency of the order of the frequency of telomere are in the lower end of the band of light coming from the sunlight and staying in the shade (in most time) people experience this infrared lights. The interference of this infrared rays with the frequency of the telomere may be one of the causes of the aging i.e., the shrinking of the skin and decaying of the normal activities of the cells of human body and finally leading to death. The frequencies of the two imaginary patterns of bases e.g., all purine and all pyrimidines are different from the original pattern with four purines and two pyrimidines. For all purines the value of the frequency is higher than that for telomeric pattern which supports for being more stable than the telomeric pattern but there may arise some problem in replicability of DNA due to the heavy attraction of the pattern GGGGGG in the first strand with CCCCCC in the second strand because GC has very high attraction between two strands. On the other hand if all purines like AAAAAA pattern exists the frequency for this pattern is the lowest among all other patterns which shows that the pattern to be unstable and this is also proved from the real data of lower energy of interaction between the bases TA or AT where one exists in one strand and the other in the other strand than that of CG or GC. Hence to continue the normal living behavior of biological molecules telomere pattern should the the most stable pattern at the replication end of DNA.

\section{ Amplitude of Vibration in Thermal Equilibrium}

\subsection{ Introduction}

Human body exists at a particular temperature. This temperature is at an average about $37^oC$. The temperature normally does not change with the change of the external
or environmental temperature. DNA molecules exist inside this temperature state. The telomeres in DNA thus exist in the thermodynamic state of
temperature $37^oC$. The telomere pattern AGGGTT has been considered as an oscillator that vibrates with some frequencies. This vibration should
create some thermodynamic motion of these molecules due to the thermal state. It is known from the well known kinetic interpretation of
temperature an atom being in the thermal state of temperature $T$ has an average translational kinetic energy which is equal to $\frac{3}{2}kT$.
Here $k$ is the Boltzmann constant that is equal to $1.38\times 10^{-23}$ joule/molecule K and $T$ is the absolute temperature of the
environment i.e., $T=273+t$, where $t$ is in $^oC$ and $T$ is in $^oK$. For one dimentional motion the average energy is one third of the previous
that is equal to $\frac{1}{2}kT$. Now any atom or molecule of mass $M$ and vibrating with the frequency should have the total energy that is
equal to $\frac{1}{2}Ma^2\omega^2$ where $a$ is the amplitude of the vibration.
In this problem it is considered for DNA telomere bases (purine and pyrimidine)
\begin{equation}
\frac{1}{2}Ma^2\omega^2=\frac{1}{2}kT
\end{equation}

The calculated values of the frequencies for telomere pattern AGGGTT and other possible random patterns have been used here. The calculation has given us the following results for the amplitudes of vibration.

\subsection{ Results and Discussions}

The following results show the amplitudes of thermal vibration for purine (Adenine and Guanine) and pyrimidine (Uracil and Thymine) in DNA.\\

TELOMERE:\\
Pyrimidine: $a =+/-0.0311 nm$\\
Purine:     $a =+/-0.0292 nm$\\

AGTGTG pattern:\\
Pyrimidine: $a =+/-0.0328 nm$\\
Purine:     $a =+/-0.0308 nm$\\

AGTGGT pattern:\\
Pyrimidine: $a =+/-0.0374 nm$\\
Purine:     $a =+/-0.0351 nm$\\

AAAAAA pattern:\\
Pyrimidine: $a =+/-0.0290 nm$\\
Purine:     $a =+/-0.0272 nm$\\

TTTTTT pattern:\\
Pyrimidine: $a =+/-0.0579 nm$\\
Purine:     $a =+/-0.0543 nm$\\

The least values for amplitudes for purine and pyrimidine are found in the telomere pattern (AGGGTT) among the patterns having two pyrimidine
and four purine base sequences in DNA. The higher the values of the thermodynamic vibrational amplitudes from the mean position of rest the
higher the possibility of the breaking of the biological structure. Hence the lowest values of the amplitudes of thermodynamic vibrational for
telomere structure support for the stability of telomere pattern from the thermodynamic point of view.

\section{ Discussion and Conclusions}

In this work some basic parameters involved in DNA, the biggest molecule, have been calculated. In doing this some approximations are considered
but these approximations do not affect the reality of the problem. Though the method of calculation is fairly simple, using sometimes somewhat
drastic approximations, it is remarkable that the results which are achieved here are either in agreement with experimental observations, or,
when that is lacking, in agreement with the idea of molecular evolution, viz. the approach to states of stability or lower energy in molecular
configurations that presumably are less susceptible to perturbations of the environment.

A possible cause of the stability of the telomeric pattern in the replication end of DNA has been suggested in this work. Before that the
interaction energies involved between the successive bases in a DNA single helix have been calculated. With these energies the values of the
force constants have been calculated for the different patterns of base combinations i.e., between the purine-purine, purine-pyrimidine and
pyrimidine-pyrimidine sequence. It is seen that the minimum energy state for the sequence purine-purine is lower than all other sequences i.e.,
purine-pyrimidine and pyrimidine-pyrimidine. In reality too it should be so, because in purine there exists one additional ring than the single
ring in pyrimidine base.

Using the values of force constants $k$'s [13] from slightly deviated portions of the minimum energy positions, the frequencies for the telomeric
pattern (AGGGTT) and other patterns have been calculated. Here the results show that the highest frequency exists in the telomeric pattern among
the different possible patterns, taking four purines and two pyrimidines. The value of the frequency for telomeric pattern is close to the
infrared band of the frequencies available in sunlight. The frequency of the telomeric pattern may have thus a resonance with the infrared rays
existing below the lower end of visible spectrum of sunlight. Human beings can not avoid exposure to direct or indirect sunlight during most of
their lifetime, and are therefore vulnerable to the effects of the infrared frequencies. The possible resonance of the telomeric frequency with
the infrared frequencies may play an important role in the aging process and other maladies such as cancer in the lives of human beings.

On the other hand this frequency for the telomeric pattern may be the cause of the thermal and mechanical stability of the pattern AGGGTT in the
replication end of DNA and therefore the pattern repeats such a large number of times. In this work the frequencies for some imaginary patterns
with all purines and with all pyrimidines in a single helix are also calculated where the results show that the frequency of the pattern with all
 purines (Adenine or Guanine) is the highest among all the patterns and the frequency for the pattern  with all pyrimidines (Thymine or Cytosine)
  is the lowest. Therefore the pattern AAAAAA or GGGGGG show most stable than other patterns. Here only a single strand of the DNA is considered.
  When both strands are considered together, of course there is an equality of the purines and pyrimidines. A problem may arise here due to the
  anomaly of attraction of these patterns with their counter strand in DNA as GC has higher attraction than AT. So if all Gs exist in the 1st coil
  then in the 2nd coil there will be all Cs. So highest binding between GC bonding may be a cause of stopping of the replication of DNA, but that
  would be impossible for continuing the living activity of human beings, or any other organism. So, for stability of the telomeric pattern
  (AGGGTT) there should be an optimization between the stability and replicability of DNA and that might be a reason why the specific telomeric
  pattern continues in human DNA at the replication end.

With the values of the frequencies of the telomeric pattern and other possible patterns of DNA bases with four purines and two pyrimidines, the
thermodynamic vibrational amplitudes have been calculated in this work. Here too telomeric pattern shows the lowest value of the amplitude of
vibration, which supports the stability of the configuration with this pattern (AGGGTT) in the thermal environment of human cells. For purines
(A or G) the amplitudes are smaller than that of pyrimidine (T or C) because purines are more massive than pyrimidines due to the presence of
extra ring in purines.

In this work some physical methods are used and some mathematical formulas have been developed for calculating the interaction energies and
frequencies of DNA molecules. Calculations have been performed with these formulas and appropriate conclusions have been drawn. In future when
more crystallographic data become available for base sequences of DNA for human genome or other living organs, the work can be extended to test
this hypothesis about the association between lower energy and preferred molecular configurations with molecular arrangements. Thus the work can
be extended to a level which can explain the most important causes of aging and other phenomena involved in living bodies, not only for human
beings but also for other living systems.\\

\underbar{Acknowledgment}

This research was partly supported by the Ministry of Science and Technology, Peoples' Republic of Bangladesh and Bangladesh University of Engineering and
Technology. The authors are grateful to Prof. Gias uddin Ahmad, Department of Physics, Bangladesh University of Engineering and Technology,  for his
valuable suggestions.\\

\bf{References}\\
\begin{scriptsize}
1.Hayflick, L., and Moorhead, P. S. (1961) Exp. Cell Res., 25, 585-621.\\
2.Hayflick, L. (1997) Biochemistry (Moscow), 62, 1380-1393 (Russ.).\\
3.Olovnikov, A. M. (1971) Dokl. Akad. Nauk SSSR, 201, 1496-1498.\\
4.Melek, M., Greene, E. C., and Skippen, D. E. (1996) Mol. Cell Biol., 16, 3437-3445.\\
5.Harley, C. B. (1991) Mutat. Res., 256, 271-282.\\
6.Kim, N. W., Piatyszek, M. A., Prowse, K. R., Harley, C. B., West, M. D., Ho, P. L. C., Coviello, G. M., Wright, W. E., Weinrich, S. L., and
Shay, J. W. (1994) Science, 266, 2011-2015.\\
7.Fengelman, M. et al.: Nature London 175, 834 (1955)\\
8.Biophysics by Walter Hoppe, Lohmann, H. Markl, H. Ziegler, ed. 2nd  p25, t2.2\\
9.G.N. Ramachandran and V. Sasisekharan, Adv. Protein Chem. 23 (1968) 283-437\\
10.G.N. Ramachandran, C. Ramakrisanan and V. Sasisekharan, J. Mol.,7 (1963) 95-99\\
11.Volkenshtein, M.V.: Molecules and Life. New York-London, Plenum Press 1970\\
12.Biophysics by Walter Hoppe, Lohmann, H. Markl, H. Ziegler, ed. 2nd p24,s2.2.4\\
13. M.Phil. thesis, submitted by Md. Ashrafuzzaman, Department of Physics, Bangladesh University of Engineering and Technology (BUET), Dhaka,
         Bangladesh,July, 2000.\\
\end{scriptsize}


 \begin{figure} [h]
\let\picnaturalsize=N
\def\picsize{14 cm}
\def\picfilename{telDNA1.eps}
\ifx\nopictures Y\else{\ifx\epsfloaded Y\else\input epsf \fi
\let\epsfloaded=Y
\centerline{\ifx\picnaturalsize N\epsfxsize \picsize\fi \epsfbox{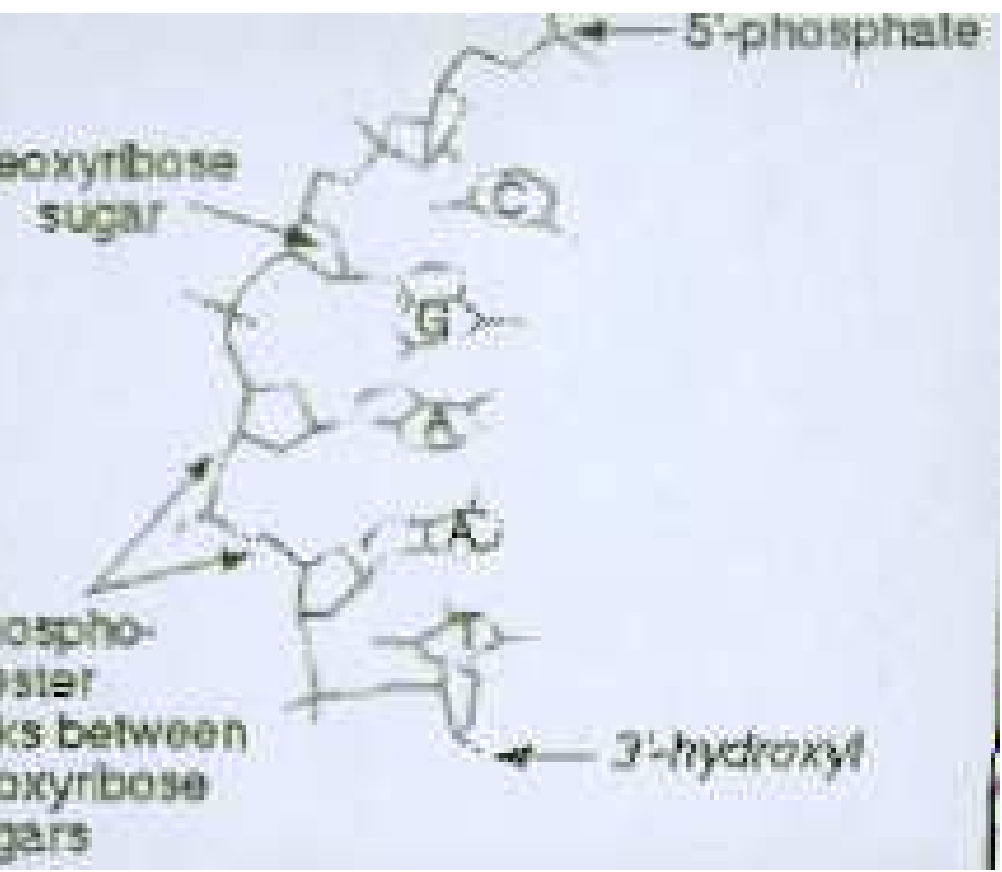}}}\fi
\caption{{\bf Sequence of bases in DNA single strand are shown. Here it is seen that the aromatic rings of the bases are parallel to each other.}}
\label{AmplitudePhi4}
\end{figure}



 \begin{figure} [h]
\let\picnaturalsize=N
\def\picsize{14 cm}
\def\picfilename{telDNA2.eps}
\ifx\nopictures Y\else{\ifx\epsfloaded Y\else\input epsf \fi
\let\epsfloaded=Y
\centerline{\ifx\picnaturalsize N\epsfxsize \picsize\fi \epsfbox{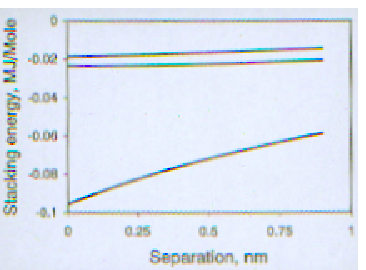}}}\fi
\caption{{\bf Stacking Energy versus Separation plot between bases in DNA. Different curves from upper to lower are for successive pyrimidine
pyrimidine, purine pyrimidine, purine purine bases in DNA.}}
\label{AmplitudePhi4}
\end{figure}



 \begin{figure} [h]
\let\picnaturalsize=N
\def\picsize{14 cm}
\def\picfilename{telDNA3.eps}
\ifx\nopictures Y\else{\ifx\epsfloaded Y\else\input epsf \fi
\let\epsfloaded=Y
\centerline{\ifx\picnaturalsize N\epsfxsize \picsize\fi \epsfbox{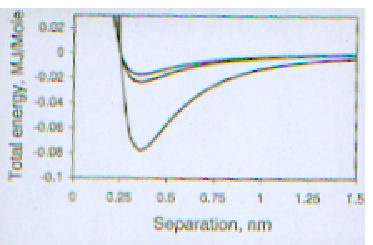}}}\fi
\caption{{\bf Total Energy (stacking+van der Waals) versus Separation plot between bases in DNA. Different curves from upper to lower are for
successive pyrimidine pyrimidine, purine pyrimidine, purine purine bases in DNA.}}
\label{AmplitudePhi4}
\end{figure}


\end{document}